# On Some New Modifications of Ridge Estimators

Yasin Asar[1a] and Aşır Genç[b]

[a]Department of Statistics, Faculty of Science, Necmettin Erbakan University, Konya, Turkey
yasar@konya.edu.tr, yasinasar@hotmail.com
[b]Department of Statistics, Faculty of Science, Selçuk University, Konya, Turkey
agenc@selcuk.edu.tr

**Abstract**

Ridge estimator is an alternative to ordinary least square estimator when there is multicollinearity problem. There are many proposed estimators in literature. In this paper, we propose new estimators which are modifications of the estimator suggested by Lawless and Wang (1976). A Monte Carlo experiment has been conducted for the comparison of the performances of the estimators. Mean squared error (MSE) is used as a performance criterion. The benefits of new estimators are illustrated using two real datasets. According to both simulation results and applications, our new estimators have better performances in the sense of MSE in most of the situations.

1. Introduction

Consider the following standard multiple linear regression model

$$Y = X\beta + \varepsilon \qquad (1.1)$$

where $Y$ is an $n \times 1$ vector of dependent variable, $X$ is an $n \times p$ design matrix consisting explanatory variables as columns where $p$ is the number of explanatory variables, $\beta$ is a $p \times 1$ vector of regression coefficients and $\varepsilon$ is an $n \times 1$ error vector distributed normally with zero mean and variance $\sigma^2$ such that $N(0, \sigma^2 I_n)$. The ordinary least squared (OLS) estimator of the coefficient vector $\beta$ is

$$\hat{\beta} = (X'X)^{-1} X'Y. \qquad (1.2)$$

---
[1] Corresponding author

In multiple linear regression, explanatory variables should be independent of each other. However in practice, there may be linear dependencies between these variables. Especially, this situation occurs in econometric data. This problem is called multicollinearity. Multicollinearity affects the regression analysis seriously. Since the basic assumption of regression analysis is not satisfied, one cannot make effective inference and prediction. Actually, this problem leads to large variance and large mean squared error (MSE).

In literature, there are some methods to solve multicollinearity. One of the most popular method is the ridge regression firstly suggested by Hoerl and Kennard (1970). They suggested using the following ridge estimator

$$\hat{\beta}_R = (X'X + kI_p)^{-1} X'Y \qquad (1.3)$$

where $k > 0$.

The MSE of the ridge estimator $\hat{\beta}_R$ is given as follows

$$MSE(\hat{\beta}_R) = \sigma^2 \sum_{j=1}^{p} \frac{\lambda_i}{(\lambda_j + k)^2} + k^2 \beta'(X'X + kI_p)^{-2} \beta \qquad (1.4)$$

where $\lambda_j$'s are the descending eigenvalues of $X'X$.

The first term of the above equation, namely, the asymptotic variance function is monotonically decreasing and the second term, namely, the squared bias function is a monotonically increasing function of the parameter $k$. Thus, there is some $k$ such that $MSE(\hat{\beta}_R)$ is less than $MSE(\hat{\beta}_{OLS}) = \sum_{j=1}^{p}(1/\lambda_j)$. However, $MSE(\hat{\beta}_R)$ depends on $\sigma^2, \beta$ and $k$ which are unknown in practice. Thus, $k$ is estimated from real data. Most of the papers on ridge regression discusses the methods of estimating the ridge parameter $k$.

In recent papers, the new suggested estimators have been compared to the one proposed by Hoerl and Kennard (1970) and each other. Many of the studies in this area suggest different estimators of ridge parameter. For detailed discussions we refer to the following studies; Hoerl et al. (1975), Lawless and Wang (1976), Saleh and Kibria (1993), Kibria (2003), Khalaf and Shukur (2005), Alkhamisi et al. (2006), Muniz and Kibria (2009), Mansson et al. (2010) and Muniz et al. (2012).

The purpose of this study is to investigate the estimation methods of ridge parameter in the literature and make a comparison between them by conducting a Monte Carlo simulation. Also, we suggest some new modifications of the estimator defined by Lawless and Wang (1976). We use MSE criterion to compare the performances of the estimators. The organization of this paper is as follows. In section 2, we give the methodology and propose some new estimators. In section 3, we present the details of Monte Carlo simulation. Moreover, we provide a real data example to illustrate the benefits of new estimators in section 4. Finally, results and discussions are given in section 5.

## 2. Methodology and Proposed Estimators

Firstly, we review the generalized ridge regression (Alkhamisi and Shukur, 2007). To do so, we write the general model (1.1) in canonical form. Assume that there exists an orthogonal matrix $D$ such that $D'X'XD = \Lambda = \text{diag}(\lambda_1, \lambda_2, \ldots, \lambda_p)$. Let us substitute $Z = XD$ and $\alpha = D'\beta$ in model (1.1), then the canonical version of (1.1) is given by the following equation

$$Y = Z\alpha + \varepsilon. \tag{2.1}$$

Thus the generalized ridge estimator is given as follows

$$\hat{\alpha}_R = (Z'Z + K)^{-1} Z'Y \tag{2.2}$$

where $K = \text{diag}(k_1, k_2, \ldots, k_p)$ such that $k_j > 0$ for each $j = 1, 2, \ldots, p$. The OLS estimator of $\alpha$ is

$$\hat{\alpha}_{OLS} = \Lambda^{-1} Z'Y. \tag{2.3}$$

The MSE of $\hat{\alpha}_R$ and $\hat{\alpha}_{OLS}$ are respectively as follows

$$MSE(\hat{\alpha}_R) = \sum_{j=1}^{p} \frac{\sigma^2 \lambda_i}{(\lambda_j + k_j)^2} + \sum_{j=1}^{p} \frac{k_j^2 \alpha_j^2}{(\lambda_j + k_j)^2} \qquad (2.4)$$

and

$$MSE(\hat{\alpha}_{OLS}) = \sigma^2 \sum_{j=1}^{p} \frac{1}{\lambda_j}. \qquad (2.5)$$

Hoerl and Kennard (1970) showed that the value of $k_j$ minimizing (2.4) is

$$k_j = \frac{\sigma^2}{\alpha_j^2} \qquad (2.6)$$

where $\sigma^2$ is the error variance and $\alpha_j$ is the $i^{th}$ element of $\alpha$. Since $\sigma^2$ and $\alpha_j^2$ are not known, they suggested to use the common unbiased estimators $\hat{\sigma}^2$ and $\hat{\alpha}$ respectively and got $\hat{k}_j = \frac{\hat{\sigma}^2}{\hat{\alpha}_j^2}$ where $\hat{\sigma}^2 = (Y'Y - \hat{\alpha}'Z'Y)/(n-p)$.

Now, we review some estimators proposed earlier and then propose our new estimators.

(1) $k_{HK} = \frac{\hat{\sigma}^2}{\hat{\alpha}_{max}^2}$

was suggested by Hoerl and Kennard (1970) where $\hat{\alpha}_{max}$ is the maximum element of $\hat{\alpha}_i$.

(2) $k_{HKB} = \frac{p\hat{\sigma}^2}{\sum_{j=1}^{p} \hat{\alpha}_j^2}$

which is the harmonic mean of $\hat{k}_j$ and suggested by Hoerl et al. (1975).

(3) $k_{LW} = \frac{p\hat{\sigma}^2}{\sum_{j=1}^{p} \lambda_j \hat{\alpha}_j^2}$

which is the harmonic mean of $k_{LW(j)} = \dfrac{\hat{\sigma}^2}{\lambda_j \hat{\alpha}_j^2}$ and proposed by Lawless and Wang (1976).

(4) $k_{AD} = \dfrac{2p\hat{\sigma}^2}{\sum_{j=1}^{p} \lambda_{max} \hat{\alpha}_j^2}$

which is the harmonic mean of $k_{AD(j)} = \dfrac{2\hat{\sigma}^2}{\lambda_{max} \hat{\alpha}_j^2}$ proposed by Dorugade (2014).

(5) $k_{KM8} = \max\left( \dfrac{1}{\sqrt{\dfrac{\lambda_{max} \hat{\sigma}^2}{(n-p)\hat{\sigma}^2 + \lambda_{max} \hat{\alpha}_j^2}}} \right)$ and

(6) $k_{KM12} = \operatorname{median}\left( \dfrac{1}{\sqrt{\dfrac{\lambda_{max} \hat{\sigma}^2}{(n-p)\hat{\sigma}^2 + \lambda_{max} \hat{\alpha}_j^2}}} \right)$

which were defined by Muniz et al. (2012).

We define our new estimators which are modifications of $k_{LW(j)} = \dfrac{\hat{\sigma}^2}{\lambda_j \hat{\alpha}_j^2}$ proposed byLawless and Wang (1976) for the generalized ridge regression. Actually, we apply the square root transformation to this individual parameter and we get $\sqrt{\dfrac{\hat{\sigma}^2}{\lambda_j \hat{\alpha}_j^2}}$ in a similar manner done in Muniz and Kibria (2009) and Mansson et al. (2010). They used this transformation successfully. After this transformation, we apply arithmetic mean, geometric mean and harmonic mean transformations and we also use maximum, minimum and median functions as used in above studies following Kibria (2003) and Muniz et al. (2012).

Thus, we get the following new estimators: Let $k_{Yj} = \sqrt{\dfrac{\hat{\sigma}^2}{\lambda_j \hat{\alpha}_j^2}}$ ,

(1) $k_{Y1} = \dfrac{1}{p} \sum_{j=1}^{p} \sqrt{\dfrac{\hat{\sigma}^2}{\lambda_j \hat{\alpha}_j^2}}$

which is the arithmetic mean of $k_{Yj}$'s.

(2) $k_{Y2} = \left( \prod_{j=1}^{p} \sqrt{\dfrac{\hat{\sigma}^2}{\lambda_j \hat{\alpha}_j^2}} \right)^{1/p}$

which is the geometric mean of $k_{Yj}$'s.

(3) $k_{Y3} = \mathrm{median}\left(k_{Yj}\right)_{j=1}^{p}$,

(4) $k_{Y4} = \mathrm{max}\left(k_{Yj}\right)_{j=1}^{p}$

which is the maximum of $k_{Yj}$'s.

(5) $k_{Y5} = \mathrm{median}\left(1/k_{Yj}\right)_{j=1}^{p}$

(6) $k_{Y6} = \mathrm{max}\left(1/k_{Yj}\right)_{j=1}^{p}$

(7) $k_{Y7} = \dfrac{1}{p} \sum_{j=1}^{p} \dfrac{1}{\sqrt{\dfrac{\hat{\sigma}^2}{\lambda_j \hat{\alpha}_j^2}}}$

which is the mean of $1/k_{Yj}$'s.

(8) $k_{Y8} = \dfrac{p}{\sum_{j=1}^{p} \sqrt{\dfrac{\lambda_j \hat{\alpha}_j^2}{\hat{\sigma}^2}}}$

which is the harmonic mean of $k_{Yj}$'s

(9) $k_{Y9} = \dfrac{p}{\sum_{j=1}^{p} \sqrt{\dfrac{\hat{\sigma}^2}{\lambda_j \hat{\alpha}_j^2}}}$

which is the harmonic mean of $1/k_{Yj}$'s

All of these estimators are compared by a Monte Carlo simulation and details of the simulation are given in section 3.

## 3. The Monte Carlo Simulation

In this section, we give the design of the Monte Carlo simulation which is conducted to compare the performances of the given estimators. In order to conduct a valuable simulation, we need to specify the effective properties of the estimators and the performance criteria. Effective factors in this simulation are the degree of correlation $\rho$ among variables, the error variance $\sigma^2$, the number of explanatory variables $p$ and the data size $n$. Also the mean squared error of the estimators has been chosen to be the performance criteria for the simulation. In order to get different degrees of multicollinearity and to generate the explanatory variables, we used the following generally used expression (see Kibria (2003)):

$$x_{ij} = \left(1 - \rho^2\right)^{1/2} z_{ij} + \rho\, z_{ip} \tag{3.1}$$

where $i = 1, 2, \ldots, n$, $j = 1, 2, \ldots, p$, $\rho^2$ is the correlation between the explanatory variables and $z_{ij}$'s are independent pseudo-random numbers following the standard normal distribution. The dependent variable $Y$ is generated by

$$Y_i = \beta_1 x_{i1} + \beta_2 x_{i2} + \ldots + \beta_p x_{ip} + \varepsilon_i \tag{3.2}$$

where $i = 1, 2, \ldots, n$ satisfying $\beta'\beta = 1$ where $\beta$ is the eigenvector corresponding to the largest eigenvalue of $X'X$ in order to get a minimized MSE due to Newhouse and Oman (1971).

We consider three different degrees of correlation, namely, $\rho = 0.90, 0.95$ and $0.99$. The sample size varies between $50$ and $200$ such that $n = 50, 100$ and $200$. The number of explanatory variables are chosen as $p = 4$ and $8$. Finally, the error variance is chosen as $\sigma^2 = 1.0$ and $5.0$. For the values of $\rho, n, p$ and $\sigma^2$, the simulation was repeated 5000 times by producing the errors in equation (3.2). For each replicate we compute $MSE(\hat{\alpha}_R)$ and $MSE(\hat{\alpha})$ by using the following equation

$$MSE(\hat{\alpha}_r) = \frac{1}{5000}\sum_{r=1}^{5000}(\hat{\alpha}_r - \alpha)'(\hat{\alpha}_r - \alpha) \qquad (3.3)$$

where $\hat{\alpha}_r$ is the estimator given in previous section at the $r^{th}$ replication.

## 4. Results and Discussions

The results of the simulation have been presented in this section. Performance of an estimator is quantified through the MSE criterion. The average MSE values (AMSE) of the estimators according to $\rho, n, p$ and $\sigma^2$ have been given in Tables 1-4. According to tables, all new proposed estimators have better performance than OLS estimator, namely, they have less AMSE than OLS estimator.

Increasing the sample size has a positive effect on the estimators, i.e, for large values of sample size, ASME values decrease as it is seen from Figure 1. It is obvious from tables that when the error variance $\sigma^2$ increases, AMSE values increases for all estimators. This result is represented for specific situations, namely, for $p=8, \rho=0.99, n=100$ in Figure 2. Moreover, an increase in the degree of correlation make a negative effect such that AMSE increase as it is observed from Figure 3.

For the case $p=4$ and $\sigma^2=1.0$, the estimator $k_{Y6}$ has the best performance among all of the estimators. However, $k_{KM8}$ is superior to other when $\rho=0.99$ and small sample sizes. For the case $p=4$ and $\sigma^2=5.0$, $k_{HKB}$ becomes the best estimator for lower degrees of correlation and $k_{Y4}$ has the lowest AMSE values for high degrees of correlation.

Moreover, for the case $p=8$ and $\sigma^2=1.0$, $k_{Y6}$ has the best estimator most of the time and $k_{Y4}$ has the lowest AMSE when $\rho=0.99$. Similarly, when $p=8$ and $\sigma^2=5.0$, although $k_{Y4}$ has the lowest AMSE most of the time, $k_{HKB}$ is superior to other estimators for lower degree of correlation and large sample sizes.

Table 1. Average MSEs of the estimator when $p = 4, \sigma^2 = 1.0$

| $\rho$ | 0.90 | | | 0.99 | | | 0.99 | | |
|---|---|---|---|---|---|---|---|---|---|
| $n$ | 50 | 100 | 200 | 50 | 100 | 200 | 50 | 100 | 200 |
| Y1 | 0.3275 | 0.2120 | 0.1171 | 0.5061 | 0.3478 | 0.2122 | 0.8369 | 0.6956 | 0.5800 |
| Y2 | 0.3301 | 0.2141 | 0.1119 | 0.5259 | 0.3662 | 0.2128 | 0.8789 | 0.7369 | 0.6196 |
| Y3 | 0.3303 | 0.2127 | 0.1110 | 0.5361 | 0.3687 | 0.2113 | 0.9210 | 0.7695 | 0.6430 |
| Y4 | 0.3476 | 0.2281 | 0.1360 | 0.5000 | 0.3505 | 0.2300 | 0.7955 | 0.6695 | 0.5593 |
| Y5 | 0.3034 | 0.1985 | 0.1072 | 0.4811 | 0.3288 | 0.1977 | 0.9808 | 0.7538 | 0.5796 |
| Y6 | **0.2333** | **0.1433** | **0.0883** | **0.3576** | **0.2278** | **0.1489** | 0.7011 | 0.5361 | **0.4110** |
| Y7 | 0.2796 | 0.1778 | 0.0990 | 0.4451 | 0.2949 | 0.1799 | 0.8821 | 0.6845 | 0.5329 |
| Y8 | 0.3551 | 0.2296 | 0.1154 | 0.5856 | 0.4081 | 0.2256 | 1.0020 | 0.8504 | 0.7113 |
| Y9 | 0.3401 | 0.2173 | 0.1114 | 0.5876 | 0.3906 | 0.2145 | 1.5461 | 1.1214 | 0.7715 |
| LW | 0.3912 | 0.2472 | 0.1187 | 0.6935 | 0.4727 | 0.2408 | 1.3995 | 1.1668 | 0.8996 |
| HK | 0.3412 | 0.2147 | 0.1118 | 0.5908 | 0.3851 | 0.2132 | 2.0763 | 1.3688 | 0.7990 |
| HKB | 0.3048 | 0.1864 | 0.1034 | 0.5017 | 0.3218 | 0.1883 | 1.6540 | 1.0655 | 0.6414 |
| AD | 0.4151 | 0.2545 | 0.1194 | 0.8508 | 0.5302 | 0.2475 | 4.5628 | 2.9008 | 1.3476 |
| KM8 | 0.3001 | 0.2008 | 0.1081 | 0.4408 | 0.3227 | 0.2009 | **0.6666** | **0.5320** | 0.4849 |
| KM12 | 0.3399 | 0.2221 | 0.1140 | 0.5476 | 0.3873 | 0.2211 | 0.8626 | 0.7748 | 0.6789 |
| OLS | 0.4191 | 0.2553 | 0.1195 | 0.8651 | 0.5332 | 0.2478 | 4.7208 | 2.9418 | 1.3526 |

Table 2. Average MSEs of the estimator when $p = 4, \sigma^2 = 5.0$

| $\rho$ | 0.90 | | | 0.99 | | | 0.99 | | |
|---|---|---|---|---|---|---|---|---|---|
| $n$ | 50 | 100 | 200 | 50 | 100 | 200 | 50 | 100 | 200 |
| Y1 | 1.3244 | 0.9091 | 0.5156 | 1.8762 | 1.4019 | 0.9107 | 2.3209 | 2.1786 | 2.0446 |
| Y2 | 1.4955 | 1.0095 | 0.5453 | 2.2153 | 1.6270 | 1.0050 | 2.8943 | 2.7179 | 2.5044 |
| Y3 | 1.5195 | 1.0207 | 0.5445 | 2.2940 | 1.6716 | 1.0129 | 3.1440 | 2.9083 | 2.6499 |
| Y4 | 1.0670 | 0.7590 | 0.4707 | **1.3751** | 1.0714 | 0.7632 | **1.5299** | **1.4527** | **1.4133** |
| Y5 | 1.5530 | 1.0046 | 0.5354 | 2.5217 | 1.7215 | 1.0004 | 4.5115 | 3.8330 | 3.0642 |
| Y6 | 1.2195 | 0.8069 | 0.4613 | 1.7448 | 1.2636 | 0.8176 | 2.1889 | 2.0900 | 1.9193 |
| Y7 | 1.5094 | 0.9768 | 0.5230 | 2.4047 | 1.6591 | 0.9758 | 3.8951 | 3.4471 | 2.8753 |
| Y8 | 1.6341 | 1.0801 | 0.5633 | 2.5611 | 1.8252 | 1.0682 | 3.7627 | 3.4242 | 3.0232 |
| Y9 | 1.7290 | 1.0977 | 0.5582 | 3.0385 | 2.0087 | 1.0835 | 7.7538 | 5.8839 | 4.0391 |
| LW | 1.7626 | 1.1505 | 0.5812 | 2.9230 | 2.0334 | 1.1311 | 5.4392 | 4.5521 | 3.6558 |
| HK | 1.1732 | 0.7721 | 0.4508 | 2.0031 | 1.3138 | 0.7729 | 8.4268 | 5.3113 | 2.8338 |
| HKB | **0.9558** | **0.6269** | **0.3874** | 1.5987 | **1.0296** | **0.6291** | 6.8312 | 4.1164 | 2.1880 |
| AD | 2.0454 | 1.2652 | 0.5963 | 4.1946 | 2.6331 | 1.2348 | 22.6858 | 14.4506 | 6.7237 |
| KM8 | 1.5875 | 1.0652 | 0.5605 | 2.2805 | 1.7238 | 1.0642 | 1.8245 | 2.1179 | 2.4347 |
| KM12 | 1.7188 | 1.1216 | 0.5701 | 2.8197 | 2.0059 | 1.1139 | 3.6712 | 3.9506 | 3.5952 |
| OLS | 2.0956 | 1.2766 | 0.5975 | 4.3257 | 2.6660 | 1.2388 | 23.6041 | 14.7092 | 6.7628 |

Table 3. Average MSEs of the estimator when $p = 8, \sigma^2 = 1.0$

| $\rho$ | 0.90 | | | 0.99 | | | 0.99 | | |
|---|---|---|---|---|---|---|---|---|---|
| $n$ | 50 | 100 | 200 | 50 | 100 | 200 | 50 | 100 | 200 |
| Y1 | 0.5670 | 0.3797 | 0.2168 | 0.7396 | 0.5627 | 0.3627 | 1.0358 | 0.9590 | 0.8317 |
| Y2 | 0.7215 | 0.4517 | 0.2330 | 0.9407 | 0.7007 | 0.4146 | 1.2997 | 1.2182 | 1.0199 |
| Y3 | 0.7554 | 0.4567 | 0.2333 | 1.0144 | 0.7271 | 0.4197 | 1.4980 | 1.3626 | 1.1021 |
| Y4 | 0.4583 | 0.3292 | 0.2274 | 0.5928 | 0.4615 | 0.3457 | **0.8610** | **0.7695** | **0.6990** |
| Y5 | 0.8515 | 0.4882 | 0.2415 | 1.1811 | 0.8097 | 0.4477 | 2.1570 | 1.7801 | 1.3128 |
| Y6 | **0.3968** | **0.2591** | **0.1600** | **0.5850** | **0.4062** | **0.2654** | 0.9830 | 0.8592 | 0.7221 |
| Y7 | 0.7139 | 0.4313 | 0.2235 | 1.0146 | 0.7028 | 0.4038 | 1.8353 | 1.5452 | 1.1684 |
| Y8 | 0.9032 | 0.5157 | 0.2499 | 1.1809 | 0.8445 | 0.4665 | 1.7150 | 1.5654 | 1.2471 |
| Y9 | 1.1448 | 0.5443 | 0.2532 | 1.7840 | 0.9923 | 0.4941 | 4.0779 | 2.9249 | 1.8220 |
| LW | 1.2359 | 0.5864 | 0.2640 | 1.6434 | 1.0502 | 0.5240 | 2.4264 | 2.1084 | 1.5833 |
| HK | 0.9926 | 0.4804 | 0.2397 | 1.7876 | 0.8400 | 0.4433 | 5.2419 | 3.2713 | 1.6524 |
| HKB | 0.5774 | 0.3047 | 0.1778 | 1.0317 | 0.5035 | 0.2985 | 3.1291 | 1.8972 | 0.9943 |
| AD | 1.8030 | 0.6249 | 0.2684 | 3.8716 | 1.3128 | 0.5614 | 11.3430 | 7.2704 | 3.0909 |
| KM8 | 0.6642 | 0.4639 | 0.2368 | 0.7984 | 0.6998 | 0.4254 | 0.8952 | 0.9351 | 0.9341 |
| KM12 | 1.1122 | 0.5537 | 0.2561 | 1.5154 | 0.9858 | 0.5022 | 2.0181 | 1.9750 | 1.5903 |
| OLS | 1.8328 | 0.6271 | 0.2687 | 3.9512 | 1.3198 | 0.5623 | 11.5169 | 7.3363 | 3.1027 |

Table 4. Average MSEs of the estimator when $p = 8, \sigma^2 = 5.0$

| $\rho$ | 0.90 | | | 0.99 | | | 0.99 | | |
|---|---|---|---|---|---|---|---|---|---|
| $n$ | 50 | 100 | 200 | 50 | 100 | 200 | 50 | 100 | 200 |
| Y1 | 2.5268 | 1.7697 | 1.0046 | 3.0642 | 2.5095 | 1.6473 | 3.5730 | 3.4507 | 3.2286 |
| Y2 | 3.3543 | 2.1772 | 1.1446 | 4.2378 | 3.2832 | 1.9962 | 5.2919 | 5.1758 | 4.4695 |
| Y3 | 3.6498 | 2.2636 | 1.1619 | 4.7105 | 3.5243 | 2.0715 | 6.1660 | 6.0272 | 4.9788 |
| Y4 | **1.3927** | 1.1165 | 0.7606 | **1.5550** | **1.4004** | 1.0912 | **1.6847** | **1.5909** | **1.5635** |
| Y5 | 4.3526 | 2.4463 | 1.2071 | 6.1623 | 4.1120 | 2.2461 | 10.9866 | 9.2113 | 6.7023 |
| Y6 | 2.5909 | 1.7378 | 0.9666 | 3.3631 | 2.6181 | 1.6582 | 3.9440 | 4.2138 | 3.7068 |
| Y7 | 4.0102 | 2.3353 | 1.1719 | 5.5955 | 3.8621 | 2.1579 | 9.2201 | 8.1761 | 6.1248 |
| Y8 | 4.0289 | 2.4214 | 1.2117 | 5.2722 | 3.8510 | 2.2028 | 7.3750 | 6.9591 | 5.5651 |
| Y9 | 5.8094 | 2.7305 | 1.2669 | 9.2274 | 5.0156 | 2.4796 | 20.9435 | 15.4628 | 9.6460 |
| LW | 4.8429 | 2.6461 | 1.2689 | 6.5049 | 4.4026 | 2.3819 | 10.3513 | 8.8126 | 6.6613 |
| HK | 3.8459 | 1.6566 | 0.9060 | 7.6554 | 3.1001 | 1.5839 | 24.4082 | 15.7781 | 7.1248 |
| HKB | 2.2390 | **1.0145** | **0.5831** | 4.3854 | 1.8160 | **0.9686** | 14.4423 | 9.1194 | 4.2350 |
| AD | 8.9600 | 3.1143 | 1.3401 | 19.2805 | 6.5465 | 2.8023 | 56.6544 | 39.9325 | 17.6938 |
| KM8 | 3.4722 | 2.4634 | 1.2369 | 4.0338 | 3.6972 | 2.2569 | 4.7571 | 4.6206 | 4.5134 |
| KM12 | 5.6448 | 2.7720 | 1.2806 | 7.8728 | 4.9735 | 2.5133 | 10.9230 | 10.4046 | 8.4176 |
| OLS | 9.1638 | 3.1356 | 1.3434 | 19.7561 | 6.5992 | 2.8117 | 57.5846 | 40.3097 | 17.7640 |

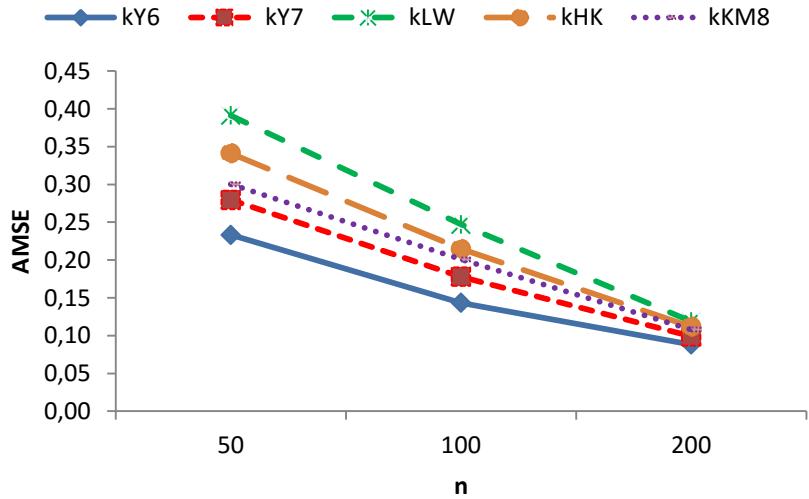

Figure 1. AMSE values of some estimators for changing values of $n$ when $p = 4, \sigma^2 = 1.0, n = 100$

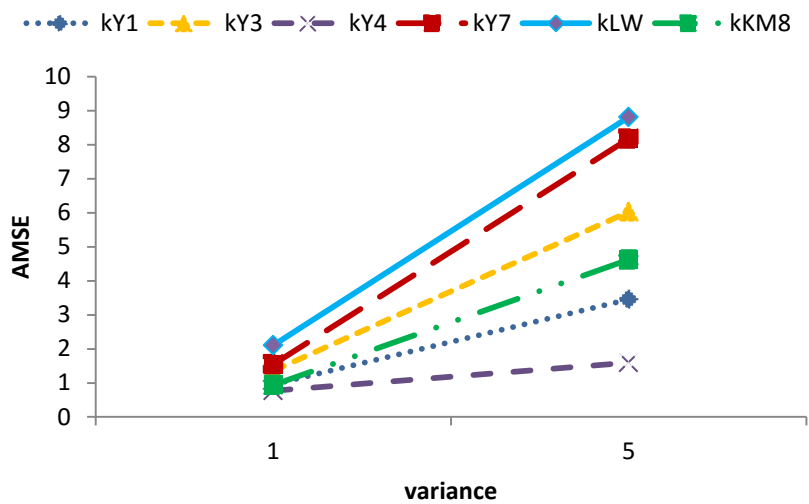

Figure 2. AMSE values of some estimators for changing values of variance when $p = 8, \rho = 0.99, n = 100$

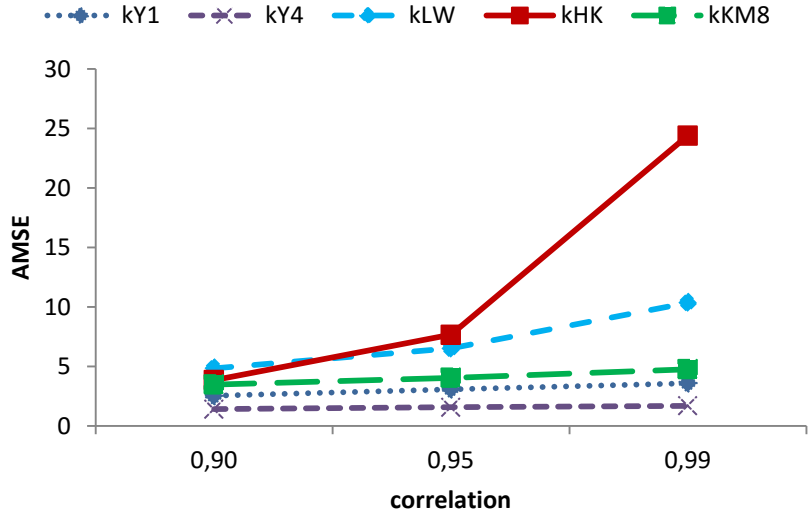

Figure 3. AMSE values of some estimators for different degrees of correlation when $p = 8, \sigma^2 = 5.0, n = 50$

## 5. Real Data Application

In order to show the performances of new estimators, we use two real data sets which are widely analyzed in the literature. First of them was studied originally by Gruber (1998). The data represents the relationship between the dependent variable Y the percentage spent by the United States and the four other independent variables the percent spent by France, that spent by West Germany, that spent by Japan, and that spent by the former Soviet Union. We firstly centered and standardized the data matrix $X$ such that $X'X$ becomes the correlation matrix. The eigenvalues of the matrix $X'X$ are obtained as 0.0202, 0.1098, 0.9122 and 2.9528. The condition number $\kappa = \max eigvalue / \min eigvalue$ of the data is approximately 146.4222 which shows moderate collinearity. The estimated theoretical MSE values of new estimators and OLS are reported in Table 5 by using equations (2.4) and (2.5). According to Table 5, $k_{Y2}$ and $k_{Y8}$ have less MSE than that of OLS.

Table 5. Estimated theoretical MSE values of new estimators and OLS

| Y1 | Y2 | Y3 | Y4 | Y5 | Y6 | Y7 | Y8 | Y9 | OLS |
|---|---|---|---|---|---|---|---|---|---|
| 0.3064 | 0.2691 | 0.3004 | 0.3611 | 0.3315 | 0.5432 | 0.4306 | 0.2100 | 0.3065 | 0.2833 |

Second real data is the widely analyzed Portland cement data which was used by Trenkler (1978) and many other authors. The data is described as follows: the effect of the composition of cement on the heat produced as it hardens. The explanatory variables are percentage, by weight, of Tricalcium Aluminate, percentage, by weight, of Tricalcium Silicate, percentage, by weight, of Tetracalcium Alumino Ferrite, percentage, by weight, of Dicalcium Silicate. The dependent variable is heat evolved in calories per gram of cement. Similarly, the data matrix is centered and standardized and the eigenvalues are as follows: 0.0016, 0.1866, 1.5761 and 2.2357. The condition number is 1376.880 which shows strict collinearity. Again, the estimated MSE values are provided in Table 6 as follows:

Table 6. Estimated theoretical MSE values of new estimators and OLS

| Y1 | Y2 | Y3 | Y4 | Y5 | Y6 | Y7 | Y8 | Y9 | OLS |
|---|---|---|---|---|---|---|---|---|---|
| 0.3226 | 0.2328 | 0.3048 | 0.4279 | 0.2971 | 0.5930 | 0.4674 | 0.1753 | 0.2156 | 1.3710 |

## 6. Conclusion

In this study, we have proposed new ridge estimators which are modifications of the estimator $k_{LW}$ defined by Lawless and Wang (1976) and studied the properties of new modified estimators for choosing ridge parameter, when there is multicollinearity between the explanatory variables. We compared the estimators proposed earlier to our new proposed estimators through a Monte Carlo simulation having 5000 replications for each combination. Average mean squared error (AMSE) has been chosen to be the evaluation criterion for the simulation. We created tables consisting of AMSE values according to different values of the sample size $n$, the degree of correlation $\rho$, the number of predictors $p$ and the variance of error terms $\sigma^2$. We have provided some figures for selected situations. According to tables and figures, we may say that our new suggestions for ridge estimators are better than the others for most of the cases. Especially $k_{Y4}$ and $k_{Y6}$ have smaller ASME values in most of the situations. Moreover, we considered two real datasets to illustrate the performances of estimators and showed the benefit of using the new estimators.

# 7. References

Alkhamisi, M. A., Khalaf, G. and Shukur, G. (2006). Some modifications for choosing ridge parameters. *Communications in Statistics—Theory and Methods, 35*(11), 2005-2020.

Alkhamisi, M. A. and Shukur, G. (2007). A Monte Carlo study of recent ridge parameters. *Communications in Statistics—Simulation and Computation®, 36*(3), 535-547.

Dorugade, A. V. (2014). New ridge parameters for ridge regression. *Journal of the Association of Arab Universities for Basic and Applied Sciences, 15*, 94-99.

Gruber, M. (1998). *Improving Efficiency by Shrinkage: The James--Stein and Ridge Regression Estimators* (Vol. 156): CRC Press.

Hoerl, A. E. and Kennard, R. W. (1970). Ridge regression: Biased estimation for nonorthogonal problems. *Technometrics, 12*(1), 55-67.

Hoerl, A. E., Kennard, R. W. and Baldwin, K. F. (1975). Ridge regression: Some simulations. *Communications in Statistics-Theory and Methods, 4*(2), 105-123.

Khalaf, G. and Shukur, G. (2005). Choosing ridge parameter for regression problems.

Kibria, B. M. G. (2003). Performance of some new ridge regression estimators. *Communications in Statistics-Simulation and Computation, 32*(2), 419-435.

Lawless, J. F. and Wang, P. (1976). A simulation study of ridge and other regression estimators. *Communications in Statistics-Theory and Methods, 5*(4).

Mansson, K., Shukur, G. and Kibria, B. G. (2010). On some ridge regression estimators: A Monte Carlo simulation study under different error variances. *Journal of Statistics, 17*(1), 1-22.

Muniz, G. and Kibria, B. G. (2009). On some ridge regression estimators: An empirical comparisons. *Communications in Statistics—Simulation and Computation®, 38*(3), 621-630.

Muniz, G., Kibria, B. G., Mansson, K. and Shukur, G. (2012). On developing ridge regression parameters: a graphical investigation. *Sort: Statistics and Operations Research Transactions, 36*(2), 115-138.

Newhouse, J. P. and Oman, S. D. (1971). An evaluation of ridge estimators.

Saleh, A. M. and Kibria, B. M. G. (1993). Performance of some new preliminary test ridge regression estimators and their properties. *Communications in Statistics-Theory and Methods, 22*(10), 2747-2764.

Trenkler, G. (1978). An iteration estimator for the linear model *COMP-STAT 1978 (Proc. Third Sympos. Comput. Statist., Leiden, 1978)* (pp. 125-131): Physika Vienna.